\documentclass[journal]{IEEEtran}

\newcommand{\vm}[1]{\mbox{\boldmath$#1$}}
\newcommand{\Prob}[1]{\Pr \left\{ #1 \right \}}

\newcommand{\ds}{\displaystyle}
\newcommand{\beq}{\begin{eqnarray}}
\newcommand{\beqq}{\begin{eqnarray*}}
\newcommand{\eeq}{\end{eqnarray}}
\newcommand{\eeqq}{\end{eqnarray*}}
\newcommand{\x}{\mbox{\boldmath$x$}}
\newcommand{\y}{\mbox{\boldmath$y$}}

\ifCLASSINFOpdf
\else
\fi

\usepackage[centertags]{amsmath}
\usepackage{amsfonts}
\usepackage{amssymb}
\usepackage{amsthm}
\usepackage{newlfont}
\usepackage{graphics}
\usepackage{mathrsfs}
\usepackage{fancyhdr}
\usepackage{graphicx}
\usepackage{multicol}
\begin{document}
%
\title{SNR Estimation in Maximum Likelihood Decoded Spatial Multiplexing}
%
%
%

\author{Oded Redlich,
        Doron Ezri,
        and Dov Wulich
\thanks{O. Redlich and D. Wulich are with the Department
of Electrical and Computer Engineering, Ben Gurion University, Beer Sheva XXX, Israel, e-mail: \{redlicho,dov\}@ee.bgu.ac.il}
\thanks{D. Ezri is with Greenair Wireless,  Raman Gan 52551, Israel, email: doron@greenairwireless.com}
\thanks{Manuscript received XXX XX, XXXX; revised XXXX XX, XXXX.}}

%
%

\markboth{Journal of XXX,~Vol.~X, No.~X, Date goes here}%
{Shell \MakeLowercase{\textit{et al.}}: Bare Demo of IEEEtran.cls for Journals}
%



\maketitle

\begin{abstract}
Link adaptation is a crucial part of many modern communications
systems, allowing the system to adapt the transmission and reception
strategies to changes in channel conditions. One of the fundamental
components of the link adaptation mechanism is signal to noise ratio
(SNR) estimation, measuring the instantaneous (mostly post
processing) SNR at the receiver. That is, the SNR at the decoder
input, which is an important metric for the prediction of decoder
performance. In linearly decoded MIMO, which is the common MIMO decoding
strategy, the post processing SNR is well defined. However, this is
not the case in optimal maximum likelihood (ML) decoding applied to
spatial multiplexing (SM). This gap is interesting since ML decoded
SM is gaining ever growing interest in recent research and practice
due to the rapid increase in computation power, and available near
optimal low complexity schemes. In this paper we close the gap and provide SNR estimation schemes
for ML decoded SM, which are based on various approximations of
the "per stream" error probability. The proposed methods are
applicable for both horizonal and vertical decoding. Moreover, we
propose a very low complexity implementation for the SNR estimation
mechanism employing the ML decoder itself with negligible overhead.
\end{abstract}


%
\IEEEpeerreviewmaketitle

\section{Introduction}
%
%
%
%
\IEEEPARstart{E}{stimating}  the signal to noise ratio (SNR) is one of the important
tasks in communications systems. The measure of SNR indicates the
quality of the channel and enables the use of link adaptation to
improve the spectral efficiency. The main idea of link adaptation is
to use the transmission parameters that yield the highest possible
bit rate. The most common parameters to be adapted are the
modulation and coding scheme. In addition some other parameters may
be adjusted for the benefit of the systems such as transmit power
levels, bandwidth usage or MIMO mode (when a MIMO scheme is
applied).

Extensive work has been performed on this area and several
approaches were introduced. In recent years, a number of new link
adaptation schemes have been proposed for different types of
wireless networks. An example for that is a
Receiver-Based Auto-Rate (RBAR) protocol based on the RTS/CTS
(Request-To-Send/Clear-To-Send) mechanism \cite{RateAdaptive}. The basic idea of RBAR
can be summarized as follows. First, the receiver estimates the
wireless channel quality using a sample of the
instantaneously-received signal strength at the end of the RTS
reception. The receiver selects the appropriate transmission rate
based on this estimate, and feeds back to the transmitter using the
CTS. Then, the transmitter responds to the receipt of the CTS by
transmitting the data packet at the rate chosen by the receiver.

In the General Packet Radio Service (GPRS) development of GSM two link adaptation schemes were proposed \cite{AlgorithmsForLA}. One is based on the estimate of the Carrier to Interference ratio (C/I),
and the other is based on the observation of the block error rate.

HIgh PErformance Radio Local Area Network type 2 (HIPERLAN/ 2)
is another wireless broadband access
system that has been specified by European Telecommunications
Standards Institute (ETSI) project BRAN (Broadband Radio Access
Network) \cite{StructureAndPerformance}.

Link adaptation is one of the key features of HIPERLAN/ 2
as it has a PHY that is very similar to 802.11a. Lin, Malmgren and Torsner
studied the system performance of link adaptation, which uses the C/I as the wireless link quality
measurement, for packet data services within HIPERLAN/2 \cite{SystemPerformanceAnalysis}.
Furthermore, Habetha and Calvo de No presented a
new algorithm for adaptive modulation and power control in a
HIPERLAN/2 network \cite{NewAdaptiveModulation}. It first assumes the maximum transmit power, and
uses the C/I observed at the receiver to determine the proper PHY
mode for the next frame transmission to meet the target packet error
rate (PER). Then, it reduces the power as much as possible while
meeting the target PER.

A different approach uses an Euclidean
distance metric to obtain channel quality information in terms of
the average signal to  noise ratio (SNR) \cite{ChannelQualityEst}. Then, a rate
adaptation scheme which uses this metric to change the modulation at
the transmitter has been described.

Link adaptation may also be applied with a user selection mechanism. In user selection, we assume that multiple users exist and their
number is larger than the number of antennas at the BS. This means
that the BS cannot receive or transmit from/to all of the
concurrently, so some selection mechanism for the formation of
groups is needed. The selection process is to be followed by link
adaptation, i.e., after the BS decides which group to transmit to, it
should decide the modulation (and coding scheme in coded systems)
for each user \cite{Optimal_Power_Allocation}, \cite{Low_complexity_user_selection}.

The IEEE 802.16 standard
introduces 2 types of carrier to interference plus
noise ratio (CINR) mechanisms \cite{IEEE80216e}. One is the physical CINR (PCINR)
which estimates the CINR or post processing CINR. The second
mechanism is the effective CINR (ECINR) which make use of the
per-tone CINR, and aims at proper weighting of the per-tone CINR to
obtain a measure for the BER \cite{Adaptive_MIMO_Transmission}. We emphasize that in MIMO
transmission-reception schemes, the post processing CINR (either
physical or effective) is the interesting measure. Thus, we focus in
the following sections on post processing SNR.

\section{Preliminaries}
\subsection{SNR in MIMO Schemes}
\begin{figure}[h]
\centering \resizebox{!}{4cm}{\includegraphics{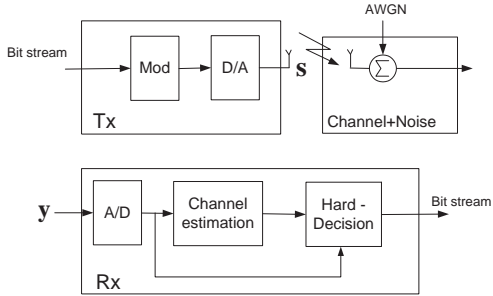}}
\caption{SISO system block diagram}
\label{Fig_SISO_BD}
\end{figure}

\begin{figure}[h]
\centering \resizebox{!}{3cm}{\includegraphics{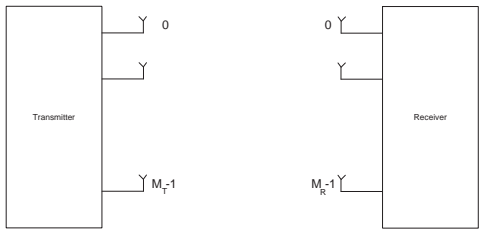}}
\caption{MIMO system block diagram (to be more detailed)}
\label{Fig_OLMIMO_BD}
\end{figure}

In order to illustrate the meaning of antenna SNR (the SNR measured at the input to the receiver) and post processing SNR, we survey linearly decoded transmission and reception schemes. We begin with the simplest single-input single-output (SISO) digital
communication system, having a single transmit (Tx) and receive (Rx)
antenna. The baseband representation of the system is illustrated in Figure \ref{Fig_SISO_BD}.
\nopagebreak
The received signal $y$ is given by
\begin{eqnarray}
y=hs+\rho v,
\label{Eq_SISO_Model}
\end{eqnarray}
where $h$ is the channel response which is assumed to be known at
the receiver (perfect channel knowledge), $s$ is the transmitted
data (e.g QAM) and  $\rho v$ is an AWGN with standard deviation
$\rho$. The instantaneous antenna SNR in this case is trivial and equals to the ratio between the signal's power to the noise's power, meaning
\begin{eqnarray}\label{Eq_SISO_Antenna_SNR}
\mbox{SNR}=\frac{|h|^2}{\rho^2}.
\end{eqnarray}
We continue with linearly decoded MIMO systems. A multiple input multiple output (MIMO) system using $M_T$ transmit
antennas and $M_R$ receive antennas is illustrated in Figure
\ref{Fig_OLMIMO_BD}.
The
received signal $\vm{y}$ is given by
\begin{eqnarray}
\vm{y}=\vm{H}_{PHY}\x+\rho\vm{v},
\label{Eq_MIMO_Model}
\end{eqnarray}
where $\vm{H}_{PHY}$ is an $\mbox{M}_R\times\mbox{M}_T$ channel matrix,
$\vm{x}$ is the $\mbox{M}_T\times1$ transmitted vector, $\vm{v}$ is a vector of i.i.d zero-mean
complex Gaussian entries with unit variance and $\rho$ is the
 noise intensity.

In spatial multiplexing (SM) for instance, independent information streams are
transmitted through the transmit antennas. Here, the
transmitted vector is $\x=\ds\frac{1}{\sqrt{M_T}}\vm{s}$, where
$\vm{s}=[s_0,s_1,\ldots,s_{M_{T}-1}]^T$ is a vector of $M_T$
independent symbols. The factor $\ds\frac{1}{\sqrt{M_T}}$ is
introduced in order to maintain unity transmission power. The
mathematical model for the received signal is
\begin{eqnarray}\label{Eq_SM_Model}
\y=\underbrace{\frac{1}{\sqrt{M_T}}\vm{H}_{PHY}}_{\vm{H}}\vm{s}+\rho\,\vm{n},
\end{eqnarray}
where $\vm{H}_{PHY}$ is the MIMO $M_R \times M_T$ channel matrix and $M_R \geq M_T$.
Applying
Zero-Forcing (ZF) detection gives
\begin{eqnarray}\label{Eq_MIMO_ZF}
\hat{\vm{s}}=\vm{H}^{\dagger}\vm{y}=\vm{s}+\rho\vm{H}^{\dagger}\vm{v}=\vm{s}+\rho\vm{Gv},
\end{eqnarray}
where $\vm{G}=\vm{H}^{\dagger}$ is the left pseudo-inverse of
$\vm{H}$. The post-processing SNR of the $i-$th stream in this case reads
\begin{eqnarray}\label{Eq_MIMO_ZF_noise_var}
\mbox{ppSNR}_i&=&\frac{1}{\mbox{var}(\hat{s}_i-s_i)}\notag \\ &=&\frac{1}{\rho^2(|G_{i,0}|^2+\ldots+|G_{i,M_R-1}|^2)}.
\end{eqnarray}
Similar results are obtained for Rx diversity and Alamouti's space-time-coding (STC) \cite{STC_Alamouti}. In these schemes linear decoding is optimal so they may be viewed as a particular case of \eqref{Eq_MIMO_ZF}.

Since in general, SM is not an orthogonal transmission scheme, linear decoding is not optimal. The optimal ML decoder in this case implies exhaustive search \cite{PaulrajBook}. In terms of post processing SNR, ML decoded SM differs from the
linear decoding methods surveyed, in the sense that at no point in
the reception process, there exists an expression for the post
processing SNR. This means that another method is to be
invoked.

\subsection{Known Approaches to SNR estimation in ML decoded SM}
In this section we consider some known approaches for SNR estimation
in ML decoded SM and related issues.
\subsubsection{SNR Estimation Based on the Capacity Formula}
This approach, that was adopted by the IEEE 802.16e standard \cite{IEEE80216e}, uses capacity computation in order to estimate the SNR. The basic idea is to use Shannon's expression for capacity to evaluate the SNR,
\begin{eqnarray}\label{Eq_SNR_Capacity_Basic}
C&=&\frac{1}{M}\log_{e}\,\det\left(\vm{I}+\frac{\vm{HH}^*}{\rho^2}\right) \notag \\ \\
\mbox{SNR Estimate}&=&e^C-1, \notag
\end{eqnarray}
where $M$ is the number of independent streams.

A natural question is the relevance of this metric to SNR. In order to answer this, we note that in the case of SISO, Rx diversity, and STC, this metric coincides with the standard post processing SNR. For example if we take the $1\times 2$ MRC scheme where $\vm{H}=\left[
                                        \begin{array}{cc}
                                          h_0 & h_1 \\
                                          \end{array}
                                          \right]^T$
we obtain
 \begin{eqnarray}
 \label{Eq_SNR_Capacity_MRC}
 C&=&
   \log\left(1+\frac{|h_0|^2+|h_1|^2}{\rho^2}\right) \notag,                                     \end{eqnarray}
which immediately implies that the estimated SNR coincides with the classical ppSNR in MRC
\begin{eqnarray}
e^{C}-1=\frac{|h_0|^2+|h_1|^2}{\rho^2}.
\end{eqnarray}

However, despite the fact that the ability  of (\ref{Eq_SNR_Capacity_Basic}) to capture the SNR in the ML decoded SM case is not established, we note that  (\ref{Eq_SNR_Capacity_Basic}) provides a single metric, and is thus inadequate for horizontal MIMO. Horizontal transmission indicates transmitting multiple separately streams over multiple antennas such that the number of streams is more than 1 (in contrary to vertical transmission that indicates transmitting a single stream over multiple antennas).

\subsubsection{SNR Estimation based on Error Probability Computation}
SNR and error probability are linked. Thus, an expression for the
error probability may be exploited to obtain an estimate for the
post processing SNR (as we will demonstrate in the next chapter).
Accordingly, we consider here the closely related problem of error
probability calculation in ML decoded SM.

The error probability in ML decoded SM does not have an analytic
solution. One of the most prominent approaches to approximate and
bound this probability is that of Paulraj and Heath
\cite{SwitchingDiversityMultiplexing}. They obtained the following
expression for the error probability
\begin{eqnarray}\label{Perr_cond_H_Paulraj}
\Prob{\mbox{error}|\vm{H}}\leq \mbox{Q}\left(\sqrt{\frac{1}{\rho^2}d^2_{\mbox{min}}(\vm{H})}\right)
\end{eqnarray}
where $d^2_{\mbox{min}}(\vm{H})$ is
\begin{eqnarray}
d^2_{\mbox{min}}(\vm{H})=\mathop{\mbox{min}\|\vm{H}(\vm{s}-\vm{c})\|^2}_{\vm{s},\vm{c}\in\mbox{QAM}^{M_T},\vm{s}\neq\vm{c}}.
\end{eqnarray}
Since the computation of $d^2_{\mbox{min}}(\vm{H})$ requires an
exhaustive search,  upper and lower bounds on the minimum Euclidean
distance were introduced.
\begin{eqnarray}\label{Eq_d_min_bounds}
\lambda_{\mbox{min}}^2(\vm{H})\frac{d_{\small{min\_QAM}}^2}{M_T}\leq d^2_{\mbox{min}}(\vm{H})\leq \lambda_{\mbox{max}}^2(\vm{H})\frac{d_{\small{min\_QAM}}^2}{M_T}
\end{eqnarray}
These bounds depend on the maximum and minimum singular values of
$\vm{H}.$  In case the condition number of $\vm{H}$ is high , these
bounds may be loose. In addition, the computation of the singular
values of $\vm{H}$, especially for high ranked $\vm{H}$, is a tough task
by itself. Moreover, like the former approach, this method results
an average estimate for the SNR over all inputs (streams), therefore
it is not suitable for the per stream SNR estimation.

The rest of the paper is organized as follows. Section II includes the derivation of a series of approximations for
the SNR in ML decoded SM based on a series of approximations for the
per stream error probability. The chapter concludes with a low
complexity implementation of the SNR estimation mechanism, based on
the ML decoder itself. In Section III we present simulation results revealing the performance
(in terms of SNR estimation error) of the various approximations in
the horizontal and vertical cases. In the vertical case we compare
our result with that of standard methods. We further show that the
QPSK based methods are valid for 16QAM and 64QAM. In Section IV we summarize the results and present topics for further
research.

\section{Proposed Method for SNR Estimation in ML Decoded SM}\label{DerivationChapter}

\subsection{A Series of Approximations for the Per-Stream SNR}\label{Section_approximation}

We base our SNR estimation method on the asymptotic evaluation of
the per stream error probability in ML decoded SM  in  high SNR. We emphasize that
in order to obtain a meaningful SNR metric, the SNR estimate should
satisfy (in QPSK) \cite{Goldsmith}
\begin{eqnarray}\label{Eq_Err_Prob_QPSK}
\mbox{p(error in stream i)}\approx e^{\ds{-\frac{\mbox{SNR}_i}{2}}}.
\end{eqnarray}We are therefore left with the problem of evaluating the per stream error probability.
The conditional probability of error given the transmitted vector $\vm{s}$ is (throughout we condition the probabilities on the channel matrix $\vm{H}$)
\begin{eqnarray}\label{Eq_Error_Prob_basic}
\Prob{\mbox{error}|\vm{s}}=\Prob{\left. \mathop{\bigcup}_{\tilde{\vm{s}}\in A(\vm{s})}\mbox{J}(\tilde{\vm{s}})<\mbox{J}(\vm{s})\right|\vm{s}},
\end{eqnarray}
where $\mbox{J}(\vm{s})=\|\vm{Y}-\vm{Hs}\|^2$, $A(\vm{s})$ is the set of all vectors corresponding to a certain \emph{type} of error, such as error in the $i-$th stream which will be represented by the set $A_i(\vm{s})$. Equation \ref{Eq_Error_Prob_basic} may be bounded using the union bound
\begin{eqnarray}\label{Eq_Union_bound}
\Prob{\mbox{error}|\vm{s}} &\leq& \sum_{\tilde{\vm{s}}\in A(\vm{s})}{\Prob{\mbox{J}(\tilde{\vm{s}})<\mbox{J}(\vm{s})|\vm{s}}},
\end{eqnarray}
which may be rewritten as
\begin{eqnarray}\label{Eq_Union_bound2}
\Prob{\mbox{error}|\vm{s}} \leq \sum_{\tilde{\vm{s}}\in A(\vm{s})}Q\left(\frac{\|\vm{H}(\tilde{\vm{s}}-\vm{s})\|}{\sqrt{2}\rho}
\right).
\end{eqnarray}
Focusing on events of error in the $i-$th stream and averaging w.r.t $\vm{s}$ gives
\begin{eqnarray}\label{Eq_Perr_1st_approx}
&&\Prob{\mbox{error in i-th stream}} \leq \\ &&\frac{1}{Q^{M_T}}\sum_{\vm{s}\in\mbox{\footnotesize{QAM}}^{M_T}}\sum_{\tilde{\vm{s}}\in
A_i(\vm{s})}{\mbox{Q}\left(\frac{\|\vm{H}(\tilde{\vm{s}}-\vm{s})\|}{\sqrt{2}\rho}\right)},\notag
\end{eqnarray}
where $Q^{-M_T}$ is a normalizing factor required due to the summation over all QAM points and transmit antennas.
Using the upper bound on the Q-function
\begin{eqnarray}\label{Eq_Q-function_approx}
\mbox{Q}(x)\leq\frac{1}{2}\exp{\left(\ds{-\frac{x^2}{2}}\right)},
\end{eqnarray}
we obtain a looser but simpler bound for the probability of the error
\begin{eqnarray}\label{Eq_Perr_2nd_approx}
&&\Prob{\mbox{error in i-th stream}}\leq \\ &&\frac{1}{2}\frac{1}{Q^{M_T}}\sum_{\vm{s}\in\mbox{\footnotesize{QAM}}^{M_T}}{\sum_{\tilde{\vm{s}}\in
A_i(\vm{s})}{e^{\ds{-\frac{\|\vm{H}(\tilde{\vm{s}}-\vm{s})\|^2}{4\rho^2}}}}}, \notag
\end{eqnarray}
which may be rewritten as
\begin{eqnarray}\label{Eq_Perr_with_e}
&&\Prob{\mbox{error in i-th stream}}\leq \\ &&\frac{1}{2}\frac{1}{Q^{M_T}}\sum_{\vm{s}\in\mbox{\footnotesize{QAM}}^{M_T}}{\sum_{\vm{e}\in
B_i(\vm{s})}{e^{\ds{-\frac{\|\vm{H}\vm{e}\|^2}{4\rho^2}}}}}, \notag
\end{eqnarray}
where $\vm{e}=\tilde{\vm{s}}-\vm{s}$  and $B_i(\vm{s})$ is the set of vectors $\vm{e}$ corresponding to $A_i(\vm{s})$.   In order to illustrate the construction of the sets $B_i(\vm{s})$ we assume that the following symbol vector was transmitted
\begin{eqnarray}\label{Eq_s_example}
\vm{s}=\ds\left[
          \begin{array}{c}
            \ds\frac{-1+j}{\sqrt{2}} \\
                                  \\
            \ds\frac{-1+j}{\sqrt{2}} \\
          \end{array}
        \right],
\end{eqnarray}
as depicted in Figure \ref{Fig_2
streams constellation map}. We focus on the probability of error in the first element, i.e., the
vectors $\vm{e}$ in which the first element is nonzero. All the
possibilities for $\vm{e}$ are illustrated in Figure \ref{Fig_2
streams constellation map} by the paths from the transmitted signal
to any possible erroneous point.
\begin{figure}[h]
\centering \resizebox{!}{4cm}{\includegraphics{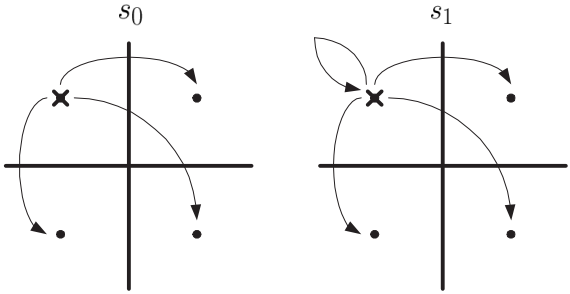}}
\caption{2 streams constellation map}
\label{Fig_2 streams constellation map}
\end{figure}
These possibilities (vectors) correspond to the following vectors
$\vm{e}$ in the set $B_0(\vm{s})$.
\begin{eqnarray}\label{Eq_B_1_full}
B_0(\vm{s})&=&\sqrt{2}\left\{
\left[
  \begin{array}{c}
    1 \\
    0 \\
  \end{array}
\right], \left[
          \begin{array}{c}
            1 \\
            1 \\
          \end{array}
        \right],\left[
          \begin{array}{c}
            1 \\
            -1 \\
          \end{array}
        \right],\left[
                 \begin{array}{c}
                   1 \\
                   1-j \\
                 \end{array}
               \right],\right. \notag \\
&&\left[
  \begin{array}{c}
    -j \\
    0 \\
  \end{array}
\right], \left[
          \begin{array}{c}
            -j \\
            1 \\
          \end{array}
        \right],\left[
          \begin{array}{c}
            -j \\
            -1 \\
          \end{array}
        \right],\left[
                 \begin{array}{c}
                   -j \\
                   1-j \\
                 \end{array}
               \right], \\
&&\left.\left[
  \begin{array}{c}
    1-j \\
    0 \\
  \end{array}
\right], \left[
          \begin{array}{c}
            1-j \\
            1 \\
          \end{array}
        \right],\left[
          \begin{array}{c}
            1-j \\
            -1 \\
          \end{array}
        \right],\left[
                 \begin{array}{c}
                   1-j \\
                   1-j \\
                 \end{array}
               \right]\right\}. \notag
\end{eqnarray}
Next, we further simplify the expression, by canceling the dependence of the sets $B_i(\vm{s})$ on the transmitted vector $\vm{s}$ and turn to unified sets $B_i$
\begin{eqnarray}
B_i=\bigcup_{\vm{s}}B_i(\vm{s}).
\end{eqnarray}
Using the unified sets, the error probability is approximated by
\begin{eqnarray}\label{Eq_Perr_with_eApprox}
\Prob{\mbox{error in i-th
stream}}\approx {\sum_{\vm{e}\in
B_i}{e^{\ds{-\frac{\|\vm{H}\vm{e}\|^2}{4\rho^2}}}}}.
\end{eqnarray}
Note that the approximation in \eqref{Eq_Perr_with_eApprox} is based on the fact that elements in the sets $B_i$ are unique, so the summation in \eqref{Eq_Perr_with_eApprox} is different than that in \eqref{Eq_Perr_with_e}. Focusing on the high SNR regime, \eqref{Eq_Perr_with_eApprox} may be well approximated by the max-log approximation as
\begin{eqnarray}\label{Eq_Perr_with_eApproxMaxLog}
\Prob{\mbox{error in i-th
stream}}\approx {e^{\ds{-\mathop{\mbox{min}}\limits_{\vm{e}\in B_i}\frac{\|\vm{H}\vm{e}\|^2}{4\rho^2}}}}.
\end{eqnarray}
This means that the sets $B_i$ may be replaced with abbreviated sets $\hat{B}_i$ that omit elements that lead identical value of $\|\vm{H}\vm{e}\|^2$, such that
\begin{eqnarray}\label{Eq_Perr_with_eApproxMaxLog2}
\Prob{\mbox{error in i-th
stream}}&\approx& {e^{\ds{-\mathop{\mbox{min}}\limits_{\vm{e}\in B_i}\frac{\|\vm{H}\vm{e}\|^2}{4\rho^2}}}}\nonumber\\
&=&{e^{\ds{-\mathop{\mbox{min}}\limits_{\vm{e}\in \hat{B}_i}\frac{\|\vm{H}\vm{e}\|^2}{4\rho^2}}}}.
\end{eqnarray}
For instance, the vectors $\left[
                                                                                   \begin{array}{cc}
                                                                                     \sqrt{2} & 0 \\
                                                                                   \end{array}
                                                                                 \right]^T$
                                                                                 and $\left[
                                                                                       \begin{array}{cc}
                                                                                         -j\sqrt{2} & 0 \\
                                                                                       \end{array}
                                                                                     \right]^T$
lead to the same cost, so one of them may be omitted. Moreover, noting that some pairs $\vm{a},\vm{b}$ of elements  in $B_i$
\begin{eqnarray}
\vm{b}=\alpha\vm{a}, |\alpha|>1
\end{eqnarray}
leads to the understanding that the cost of $\vm{b}$ is $\alpha^2$ the cost of $\vm{a}$, so $\vm{b}$ may also be omitted from the abbreviated set $\hat{B}_i$. An example to such a pair is
\begin{eqnarray}\label{Eq_redundent_e}
\ds\left[
          \begin{array}{c}
            \sqrt{2} \\
                     \\
            \sqrt{2}\\
          \end{array}
        \right],
\end{eqnarray}
and
\begin{eqnarray}\label{Eq_redundent_e}
\ds\left[
          \begin{array}{c}
            \sqrt{2}+j\sqrt{2} \\
                     \\
            \sqrt{2}+j\sqrt{2}\\
          \end{array}
        \right].
\end{eqnarray}
These arguments lead to the abbreviated set $\hat{B}_0$
\begin{eqnarray}\label{Eq_B_0_redundant}
\hat{B}_0=\sqrt{2}\left\{
\left[
  \begin{array}{c}
    1 \\
    0 \\
  \end{array}
\right], \left[
          \begin{array}{c}
            1 \\
            1 \\
          \end{array}
        \right],\left[
          \begin{array}{c}
            1 \\
            j \\
          \end{array}
        \right],\left[
                 \begin{array}{c}
                   1 \\
                   -1 \\
                 \end{array}
               \right],\left[
                 \begin{array}{c}
                   1 \\
                   -j \\
                 \end{array}
               \right],\right. \notag \\ \notag \\
               \left[
                        \begin{array}{c}
                          1 \\
                          1-j \\
                        \end{array}
                      \right],\left[
                 \begin{array}{c}
                   1 \\
                   1+j \\
                 \end{array}
               \right],\left[
                        \begin{array}{c}
                          1\\
                          -1+j \\
                        \end{array}
                      \right],\left[
                        \begin{array}{c}
                          1\\
                          -1-j \\
                        \end{array}
                      \right], \\ \notag \\
               \left.\left[
                        \begin{array}{c}
                          1+j \\
                          1 \\
                        \end{array}
                      \right],\left[
                 \begin{array}{c}
                   1+j \\
                   -1 \\
                 \end{array}
               \right],\left[
                        \begin{array}{c}
                          1+j \\
                          j\\
                        \end{array}
                      \right],\left[
                        \begin{array}{c}
                          1+j \\
                          -j\\
                        \end{array}
                      \right]\right\}. \notag \\ \notag
\end{eqnarray}
An interesting outcome resulting from the abbreviation procedure is that the sets $\hat{B}_i, \, \hat{B}_j,\, i\neq j$ differ only in a single vector, denoted here as the first. The first vector $\vm{b}_i$ in the $i-$th set, refers to the error vector in which only the $i-$th element is nonzero. This implies that the sets $\hat{B}_i$ may be written as
\begin{eqnarray}
\hat{B}_i=\{ \vm{b}_i, C\},
\end{eqnarray}
where the set $C$ is the intersection of all sets $\hat{B}_i$,
\begin{eqnarray}
C=\bigcap^i \hat{B}_i.
\end{eqnarray}

\subsection{Low Complexity Implementation}
\begin{figure}[h]
\centering \resizebox{!}{4cm}{\includegraphics{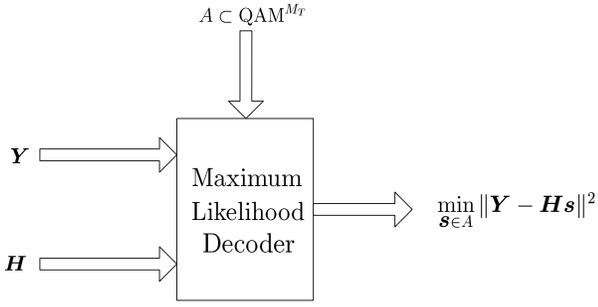}}
\caption{Maximum likelihood decoder}
\label{Fig_ML_Decoder}
\end{figure}

The structure of \eqref{Eq_Perr_with_eApproxMaxLog2} implies that the SNR may be evaluated using the ML decoder. Such a decoder is depicted in Fig. \ref{Fig_ML_Decoder}.
The ML decoder uses the inputs $\vm{Y}$ and $\vm{H}$ and performs a search over all vectors from a predefined set $\mathcal{A}$ (usually $\mathcal{A} \subset \mbox{QAM}^{M_T}$). The output of the decoder is $\mathop{\mbox{min}}\limits_{\vm{s}\in A_{ml}}\|\vm{Y}-\vm{Hs}\|^2$. We notice that with setting $\vm{Y}=0$ and searching over $\hat{B}_i$ instead of $\mathcal{A}$ we can evaluate the SNR introduced in \eqref{Eq_Perr_with_eApproxMaxLog2}. Since the sets  $\hat{B}_i$ are significantly smaller that $\mathcal{A}$ and since the SNR calculation process is done once for many information bits, the SNR estimation task embodies a small to negligible fraction of the ML decoder resources.

To set ideas straight, we assume we have a 10 symbol QPSK allocation
of 2 antenna SM. The data decoding process invokes the ML decoder
twice for each transmitted bit. Each search is over
$\frac{1}{2}\mbox{QPSK}^2$ or 8 constellation points. All in all we
have 80 searches over 8 constellation points. In this allocation a
single SNR calculation is needed, so we have the ML decoder
invoked twice (once for each stream) to search a set of 9
elements. This means that in this case the extra overhead introduced
by the SNR calculation is $\ds\frac{2}{80}=2.5\%$.

We note that in many receivers the optimal ML decoder considered here is replaced with lower complexity near optimal decoders (\cite{Low_complexity_scalable},\cite{Low_complexity_suboptimal}). In this case, the SNR estimation
methods may be modify to accommodate near optimal ML decoders.

\section{Simulation Results}\label{Chapter_Simulation_Results}
In this chapter we present simulation studies investigating the
performance of the series of error probability approximations
presented in the previous chapter and the corresponding estimated
SNR.  The simulation results reveal the following interesting
virtues of the proposed SNR estimation methods.
\begin{itemize}
\item The performance of the proposed methods are consistent with
the level of approximation. This means that the most accurate
approximation \eqref{Eq_Perr_1st_approx} relying on solely on the
union bound is superior to all other, while the simplest
\eqref{Eq_Perr_with_eApproxMaxLog2} is inferior to all others.
\item The approximations introduce small performance degradation, so the performance of all methods is similar.
\item In case of vertical encoding (where other methods are
applicable) the proposed methods slightly outperform the existing
method. This serves as a good "sanity check" for the methods.
\item SNR should not depend on the modulation employed. Thus, we examine the application of a QPSK based SNR estimation mechanism also for higher modulations. We show that the QPSK based mechanism gives plausible results also for 16QAM.
\end{itemize}

\subsection{The Simulation Setup}

In order to investigate the performance of the proposed methods, we
used the following procedure.

\begin{enumerate}
\item Randomly generate 2000 i.i.d Rayleigh distributed $2\times 2$ matrices $\vm{H}$.
\begin{enumerate}
\item Using each matrix transmit $10^6$ vectors
$\vm{s}\in\mbox{QAM}^2$, and decode the vectors $\vm{s}$ from
\begin{eqnarray}
\y=\frac{1}{\sqrt{2}}\vm{Hs}+\rho\,\vm{n},
\end{eqnarray}
using the ML decoder.
\item Compute the empiric SER in each stream.
\item Compute the empiric post processing SNR implied by the per
stream SER according to
\begin{eqnarray}
\mbox{SNR}_i=-2\log\mbox{SER}_i.
\end{eqnarray}
\item Compute the estimated SNR for QPSK at each stream according to the
three approximations \eqref{Eq_Perr_1st_approx}, \eqref{Eq_Perr_with_eApprox} and \eqref{Eq_Perr_with_eApproxMaxLog2} (denoted as "Union Bound", "Full-sum approx" and "Max-log approx" respectively in the graphs).
\item Compute the empiric joint SER for both streams.
\item Compute the empiric post processing SNR implied by the joint
SER.
\item Compute the joint estimated SNR for both streams based on the
per stream estimated SNR through the equation
\begin{eqnarray}\label{Eq_SNR_vert}
\hat{\mbox{SNR}}_{vert}[\mbox{dB}]=\frac{1}{2}\left(
\hat{\mbox{SNR}}_1[\mbox{dB}]+\hat{\mbox{SNR}}_2[\mbox{dB}]\right),
\end{eqnarray}
where $\hat{\mbox{SNR}}_i$ is the estimated SNR in the $i-$th stream
based on \eqref{Eq_Perr_1st_approx}, \eqref{Eq_Perr_with_eApprox} and
\eqref{Eq_Perr_with_eApproxMaxLog2}.
\item Compute the capacity based joint SNR, $\hat{\mbox{SNR}}_{cap}$, according to the capacity method \eqref{Eq_SNR_Capacity_Basic} (denoted as "Capacity" in the graphs).
\end{enumerate}
\end{enumerate}

\subsection{Results for Horizontal QPSK}

\begin{figure}[htp]
\centering
\makebox[5cm]{\includegraphics[width=7cm]{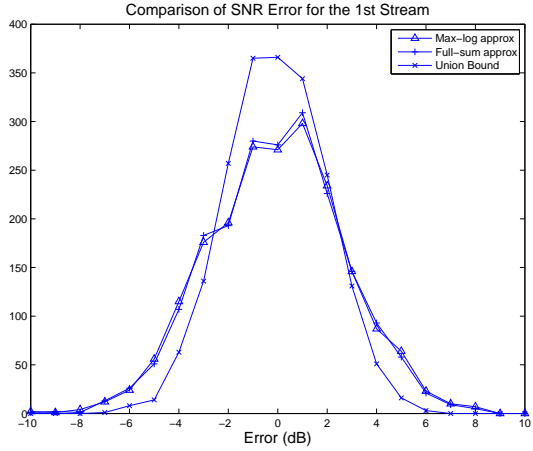}}
  \caption{Distribution of the SNR estimation error respective to stream 0.}\label{Fig_SNR_Error_hist_stream0}
\end{figure}

The simulation results show that the standard deviation of the error is approximately 2-3dB for each stream using both the full-sum and max-log approximations. Moreover, the standard deviation of the error when using the union bound approximation is some 1dB, which means only 1dB better than the former approximations. These differences are well shown in Figure \ref{Fig_SNR_Error_hist_stream0}. We note that each graph was compensated with its mean error so it is symmetric around zero. It is easy to see that the max-log approximation based SNR and the full-sum approximation based SNR are very similar for stream 0. (The same holds for stream 1, too). Hence we may conclude that the max-log approximation introduces negligible performance degradation.

\subsubsection{Vertical QPSK}

\begin{figure}[htp]
  \makebox[9cm]{\includegraphics[width=7cm]{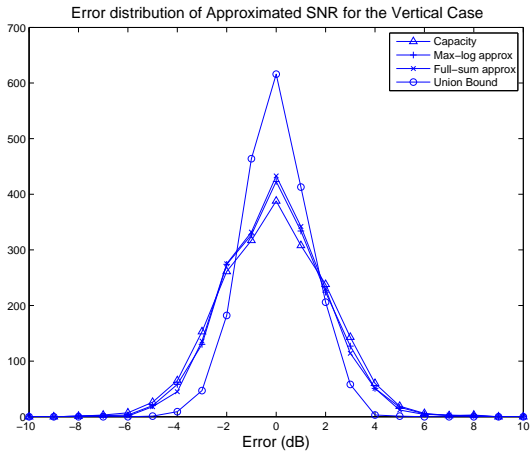}}
  \caption{Distribution of the SNR estimation error for the vertical case}\label{Fig_SNR_Error_hist_Vertical}
\end{figure}

In order to expand the algorithm to support a single SNR we defined the SNR to be the logarithmic average of the 2 per stream SNRs
\begin{eqnarray}\label{Eq_Vertical_SNR_Average}
\mbox{SNR}_{Vertical} = \frac{1}{2}(\mbox{SNR}_0+\mbox{SNR}_1),
\end{eqnarray}
for the union bound, full-sum and max-log approximations. Fig. \ref{Fig_SNR_Error_hist_Vertical} shows the error distribution of the vertical SNR calculation along with the capacity-based SNR as described in \eqref{Eq_SNR_Capacity_Basic}. The graph shows that there is a slight advantage for the new algorithm (for both the full-sum and max-log approximations, which similarly to the horizontal case have very close performance) over the capacity based SNR in terms of error standard deviation. As expected, the union bound SNR is superior to all three other calculations, having a standard deviation of approximately 1dB.

\subsection{16QAM and 64QAM}
As mentioned, the SNR should not be constellation dependent. Therefore we examine the new method when 16QAM and 64QAM constellations are applied. The only modification needed is the calculation of the reference SNR which for QPSK was introduced in (\ref{Eq_Err_Prob_QPSK}). In the case of 16QAM the error probability is given by
\begin{eqnarray}\label{Eq_SNR_BER_16QAM}
\mbox{p(error in stream i)}=e^{\ds{-\frac{\mbox{SNR}_i}{10}}},
\end{eqnarray}
and in the case of 64QAM it reads
\begin{eqnarray}\label{Eq_SNR_BER_64QAM}
\mbox{p(error in stream i)}=e^{\ds{-\frac{\mbox{SNR}_i}{42}}},
\end{eqnarray}
The values in the exponent denominators are obtained from the normalized constellation construction as depicted in the IEEE 802.16 standard \cite{IEEE80216e}.

\subsubsection{Preliminary verification}
\begin{figure}[htp]
\makebox[9cm]{\includegraphics[width=7cm]{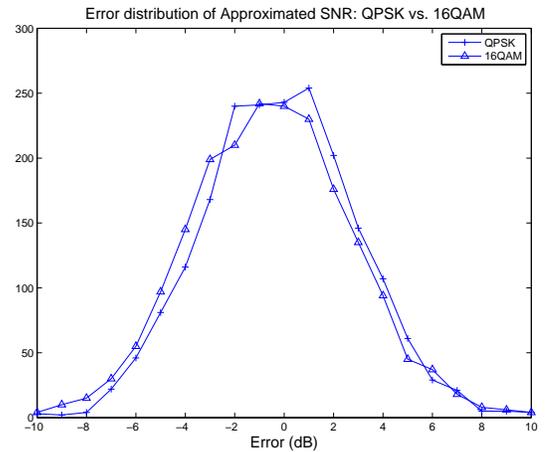}}
  \caption{Distribution of the SNR estimation for QPSK and 16QAM}\label{Fig_Err_dist_QPSK_vs_16QAM}
\end{figure}

A straight forward extension of the ideas demonstrated in Section \ref{Section_approximation} is to develop a 16QAM based SNR estimation mechanism for 16QAM transmission and a 64QAM mechanism for 64QAM transmission. The performance of a 16QAM based mechanism for 16QAM transmission is given in Fig. \ref{Fig_Err_dist_QPSK_vs_16QAM}. Simulation results show that the performance in the 16QAM case is similar to that in QPSK (the standard deviation is approximately 2dB) .

Bearing in mind that SNR should be a modulation independent measure, we suggest to employ a QPSK based SNR estimation mechanism for other modulations as 16QAM and 64QAM. Obviously, this approach leads to a more efficient mechanism since it implies a single mechanism for all modulations and more important, it makes use of the smaller search spaces respective to QPSK (the 16QAM based set, which is not introduced in this work, consists of 50 entries compared to only 13 entries in the QPSK based set, as given in (\ref{Eq_B_0_redundant}).  The performance of a QPSK based mechanism for 16QAM transmission is given in Fig. \ref{Fig_Err_dist_QPSK_vs_16QAM}. Note that the performance of the QPSK based mechanism is slightly inferior to the 16QAM based mechanism, however, this degradations seems negligible.

\subsubsection{Per Stream SNR for 16QAM}
\begin{figure}[htp]
\makebox[9cm]{\includegraphics[width=7cm]{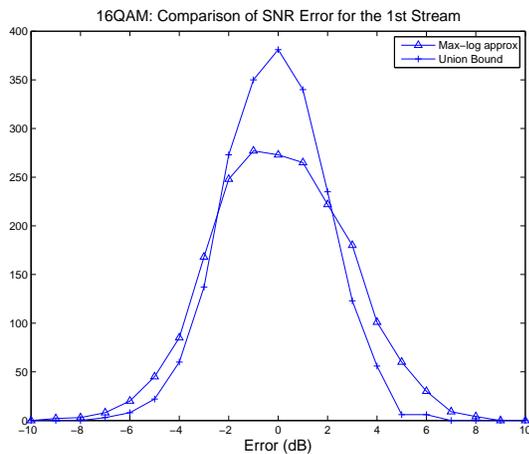}}
  \caption{Distribution of the SNR estimation error of stream 0 for 16QAM}\label{Fig_SNR_Error_hist_stream0_16QAM}
\end{figure}

Simulation report indicated a standard deviation of ~2.2dB for the union bound SNR calculation and ~3dB for the approximated SNR calculation For the horizontal case. As for the vertical case, similarly to QPSK, a small degradation in means of error standard deviation ($\sim 0.1$dB) was achieved using the approximated SNR over the traditional capacity based SNR. Figure \ref{Fig_SNR_Error_hist_stream0_16QAM} shows the distributions of the SNR error for stream 0.
\section{Discussion and Conclusions}\label{Chapter_Summary}

In this paper, we develop methods for estimating the post processing
SNR in ML decoded SM. Our results include a series of SNR estimation
methods based on various approximations for the per stream error
probability in ML decoded SM. We propose a very low complexity
implementation of the SNR estimation method, based on the ML decoder
itself with negligible overhead over the routine employment of the
decoder for data detection. We show that QPSK based algorithm provides
plausible performance also for the higher modulations, so that a single
SNR estimation mechanism is required for link adaptation.
Per stream SNR estimation for ML decoded SM may play an important role in UL and DL SDMA schemes where each stream corresponds to a different user, hence per stream SNR estimation is required \cite{Optimality_of_ZFBF}, \cite{ZF_on_SDMA}, \cite{On_the_Optimality}



%

\appendices

\ifCLASSOPTIONcaptionsoff
  \newpage
\fi



\bibliographystyle{IEEEtran}
\bibliography{OdedPaper}
%

%

%
%
%




\end{document}